\documentclass[prb,twocolumn,superscriptaddress,showpacs]{revtex4}
\usepackage{graphicx}
\usepackage{times}
\usepackage{dcolumn}
\usepackage{natbib}
\begin{document}
\title{Optimized Gaussian exponents for Goedecker-Teter-Hutter pseudopotentials}
\author{Eeuwe S.\ Zijlstra}
\email{Zijlstra@physik.uni-kassel.de}
\author{Nils Huntemann}
\affiliation{Theoretische Physik, Universit\"at Kassel, Heinrich-Plett-Str.\ 40, 34132 Kassel, Germany}
\author{Alan Kalitsov}
\affiliation{Theoretische Physik, Universit\"at Kassel, Heinrich-Plett-Str.\ 40, 34132 Kassel, Germany}
\affiliation{Solid State Theory, Institute of Physics, Lund University, S\"olvegatan 14A, S-22362 Lund, Sweden}
\author{Martin E.\ Garcia}
\affiliation{Theoretische Physik, Universit\"at Kassel, Heinrich-Plett-Str.\ 40, 34132 Kassel, Germany}
\author{Ulf von Barth}
\affiliation{Solid State Theory, Institute of Physics, Lund University, S\"olvegatan 14A, S-22362 Lund, Sweden}
\begin{abstract}
We have optimized the exponents of Gaussian \textit{s} and \textit{p} basis functions for the elements H, B--F, and Al--Cl
using the pseudopotentials of Goedecker, Teter, and Hutter [Phys. Rev. B \textbf{54}, 1703 (1996)]
by minimizing the total energy of dimers.
We found that this procedure causes the Gaussian to be somewhat more localized than the usual procedure, where
the exponents are optimized for atoms. 
We further found that three exponents, equal for \textit{s} and \textit{p} orbitals,
are sufficient to reasonably describe the electronic structure of all elements that we have studied.
For Li and Be results are presented for pseudopotentials of Hartwigsen \textit{et al.} 
[Phys. Rev. B \textbf{58}, 3641 (1998)]. 
We expect that our exponents will be useful for density functional theory studies where speed is important.
\end{abstract}
\maketitle

\section{Introduction}

Density functional theory \cite{Hohenberg64,Kohn65} (DFT) has become a standard tool for
electronic structure calculations in condensed-matter physics.
Using computer clusters, nowadays even solids with several hundreds of atoms per unit cell 
can routinely be studied.
For molecular dynamics simulations of large systems, however, existing codes are still
too slow to be able to perform large numbers of time steps within a reasonable amount of time.
To overcome this problem, faster DFT codes are needed.
To this purpose two approximations are very useful.
First of all, pseudopotentials eliminate the need to take into account core electrons
and simplify the computation for the valence electrons by being softer than the
original all-electron potentials, thus reducing the number of basis functions that have to be
included in practical calculations.
Secondly, the use of localized (atom-centered) basis states makes it possible to compute many
quantities, such as, the electronic density, by means of procedures that scale linearly
with the number of atoms in the unit cell [so-called order($N$) methods].
Unfortunately, although many different pseudopotentials have been published throughout the years, 
the problem of finding the best possible local basis functions for these pseudopotentials is only rarely addressed.
One reason is that the form of such optimized basis states depends on the pseudopotentials used
and is thus not universal.
Another reason is perhaps that local basis states, especially Gaussians, are mainly used by quantum chemists,
who are often interested in getting the most accurate results possible, which arguably
speaks against approximating the all-electron potential by pseudopotentials.
We would here like to stress that our goal is to perform calculations on large systems, trading
accuracy for speed, and to compete with tight-binding methods predominantly used for such systems. 
\cite{footnote1,Jeschke01,Jeschke02}

Toward our proposed goal
we present, in this paper, optimized Gaussian exponents for the pseudopotentials of Goedecker,
Teter, and Hutter. \cite{Goedecker96}
We chose Gaussians as basis functions in part because a lot of literature exists 
that describes how to calculate Hamiltonian matrix elements in an efficient, analytical way,
and in part because the tails of Gaussian functions decay very rapidly.
The latter property can be exploited to speed up the ensuing computer code.
We have selected the pseudopotentials of Goedecker, Teter, and Hutter, \cite{Goedecker96}
for the following reasons:
(1) They have an analytical form, which is characterized by only a few parameters.
(2) Their form has been optimized for efficient real-space DFT computations.
(3) They are available for the first two rows of the periodic table.
(4) pseudopotentials of a slightly generalized form are available for elements throughout the 
periodic table [See Hartwigsen \textit{et al.} (Ref.\ \onlinecite{Hartwigsen98})].
(5) Their validity has been thoroughly tested, \cite{Goedecker96} among other ways by comparing bond lengths
of 34 small molecules with all-electron DFT results.
(6) They can be used in \textsc{abinit}, \cite{abinit} a freely available DFT program, which uses a plane wave basis.
We used \textsc{abinit} to obtain highly accurate reference results for dimers of the elements that we studied
as well as for a number of other small molecules.
The remainder of this paper is organized as follows.
In Sec.\ \ref{sec_boron} we describe our DFT calculations.
Our optimized exponents are presented in Sec.\ \ref{sec_results}.
We conclude in Sec.\ \ref{sec_conclusion}.

\section{\label{sec_boron}Computational procedure}

For our DFT calculations we wrote a program following the work of Obara and Saika, \cite{Obara86}
which gives recursive expressions for overlap, kinetic, and Hartree matrix elements of
\textit{s}- and \textit{p}-type orbitals, starting from the much easier expressions for \textit{s}-type orbitals.
Corresponding expressions for the Goedecker-Teter-Hutter pseudopotentials were straightforwardly derived.
The exchange and correlation potential (we used the local density approximation of Perdew and Wang \cite{Perdew92})
was calculated on a fine mesh of grid points, without further approximations.
For the Hartree matrix elements the exact four-center integrals were calculated.
No intermediate fitting procedure was used for the charge density.
In the selfconsistent loop
the density matrix, describing the electronic charge density, was mixed using Broyden mixing. \cite{Broyden65}
Electronic occupation numbers were calculated according to the Fermi-Dirac statistics with
an electronic temperature of $1$ mHa.
Because our main interest is in optimizing exponents for speed, 
we limited ourselves to Gaussian basis states with \textit{s} and
\textit{p} character only, an approximation which should work best for light elements.

As mentioned above, reference results were obtained with the computer program \textsc{abinit}. \cite{abinit}
This program uses periodic boundary conditions and a plane wave basis.
For the \textsc{abinit} results presented below, a careful convergence study was made 
with respect to the box size and the number of plane waves included to ensure that 
the total free energies obtained were converged to within $1$ mHa or better.

\section{\label{sec_results}Results}

Usually Gaussian exponents are optimized for atoms, which may not give optimal
results for molecules or solids.
To improve on this situation we used dimers for our optimizations.
We found that the Gaussian functions became somewhat more localized as compared to the atomic optimizations, which
is favorable if speed is important (fewer matrix elements need to be calculated).
For Si we additionally optimized the exponents for three atoms arranged on the corners of a triangle and for four
atoms located on the vertices of a square. 
For these latter optimizations we found no systematic change in the exponents.
Therefore, we decided to optimize our Gaussian basis functions using dimers.

In our optimization procedure the total free energy of the dimers was minimized including different 
numbers of \textit{s} and \textit{p} orbitals in the basis.
The following configurations were tested: 1\textit{sp}, 1\textit{s}1\textit{p}, 1\textit{s}2\textit{p},
2\textit{sp}, 2\textit{s}1\textit{p}, 2\textit{s}2\textit{p}, 2\textit{s}3\textit{p}, 
3\textit{sp}, 3\textit{s}1\textit{p}, 3\textit{s}2\textit{p}, 3\textit{s}3\textit{p},
4\textit{sp}, 4\textit{s}3\textit{p}, 4\textit{s}4\textit{p},
where $n$\textit{s}$m$\textit{p} means that $n$ \textit{s}-type and
$m$ \textit{p}-type Gaussians were used, and
$n$\textit{sp} indicates that every one of $n$ exponents were used for \textit{s} and \textit{p}
Gaussians.
The possibility of equal \textit{s}- and \textit{p}-exponents was included,
because it is computationally advantageous (see, for example, Ref.\ \onlinecite{Obara86}).
After each optimization, we computed the binding energy of the dimer
for a wide range of atomic separations.

\begin{figure*}
  \includegraphics[angle=-90,width=7.4cm]{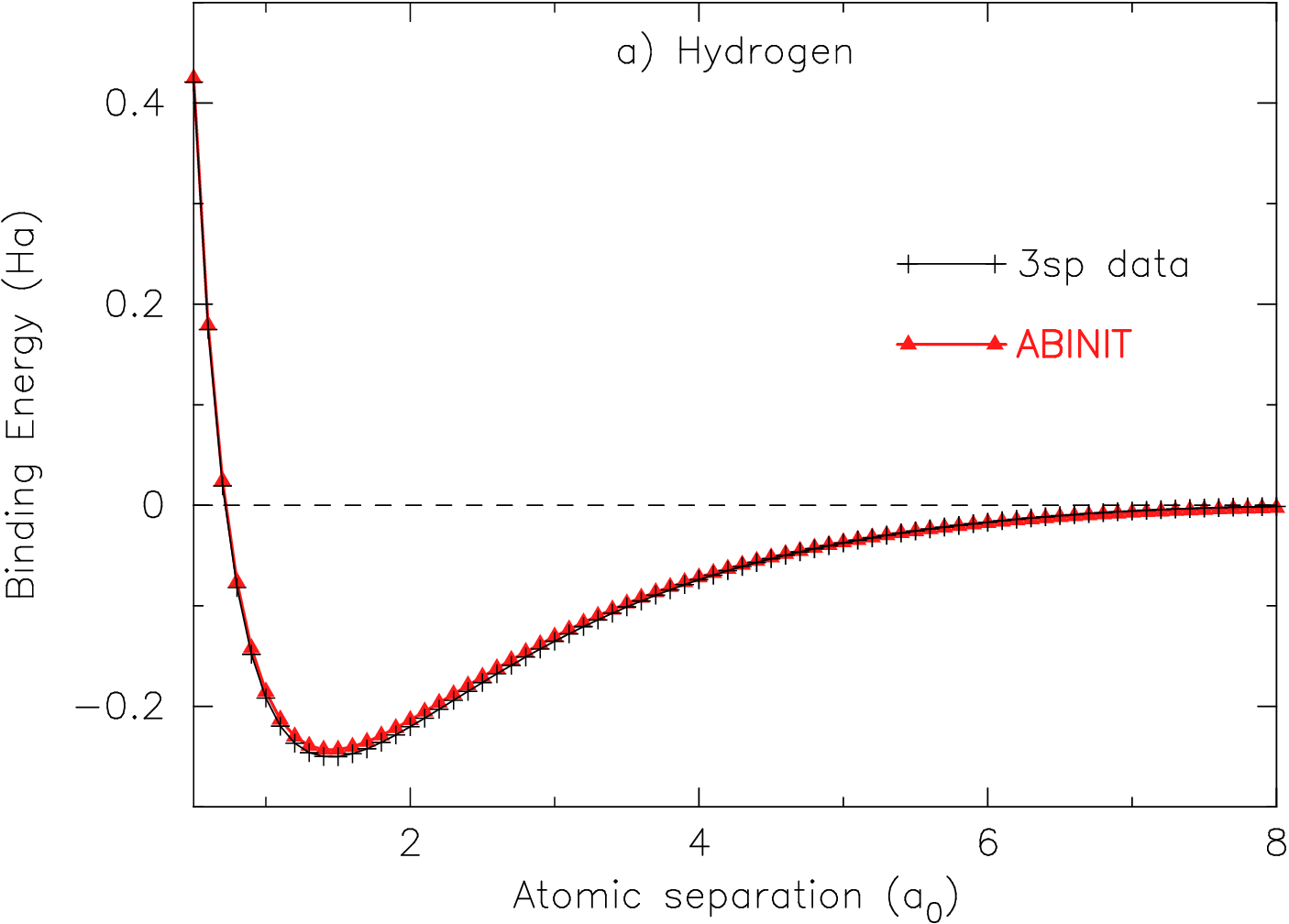}
  \includegraphics[angle=-90,width=7.4cm]{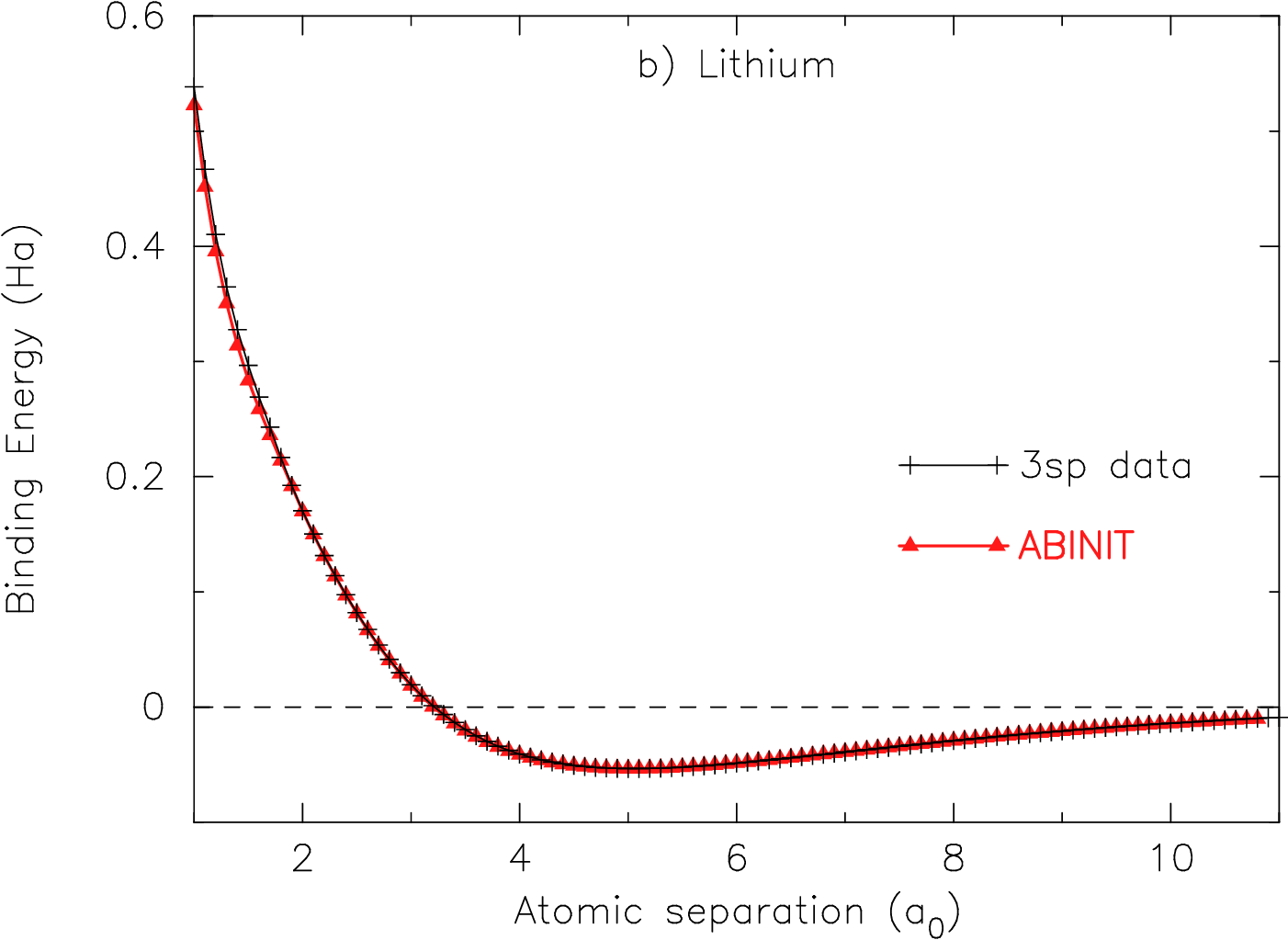}
  \includegraphics[angle=-90,width=7.4cm]{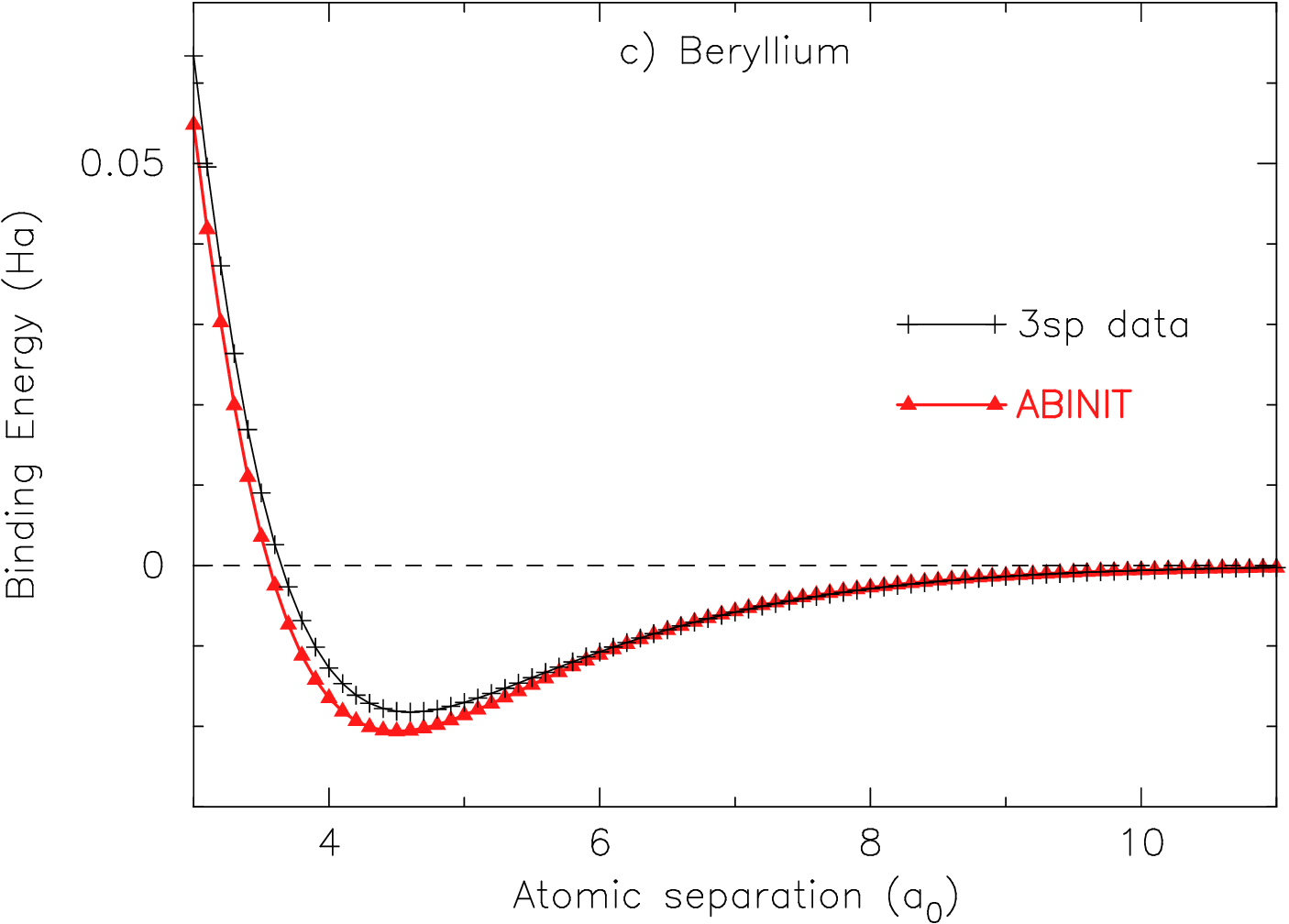}
  \includegraphics[angle=-90,width=7.4cm]{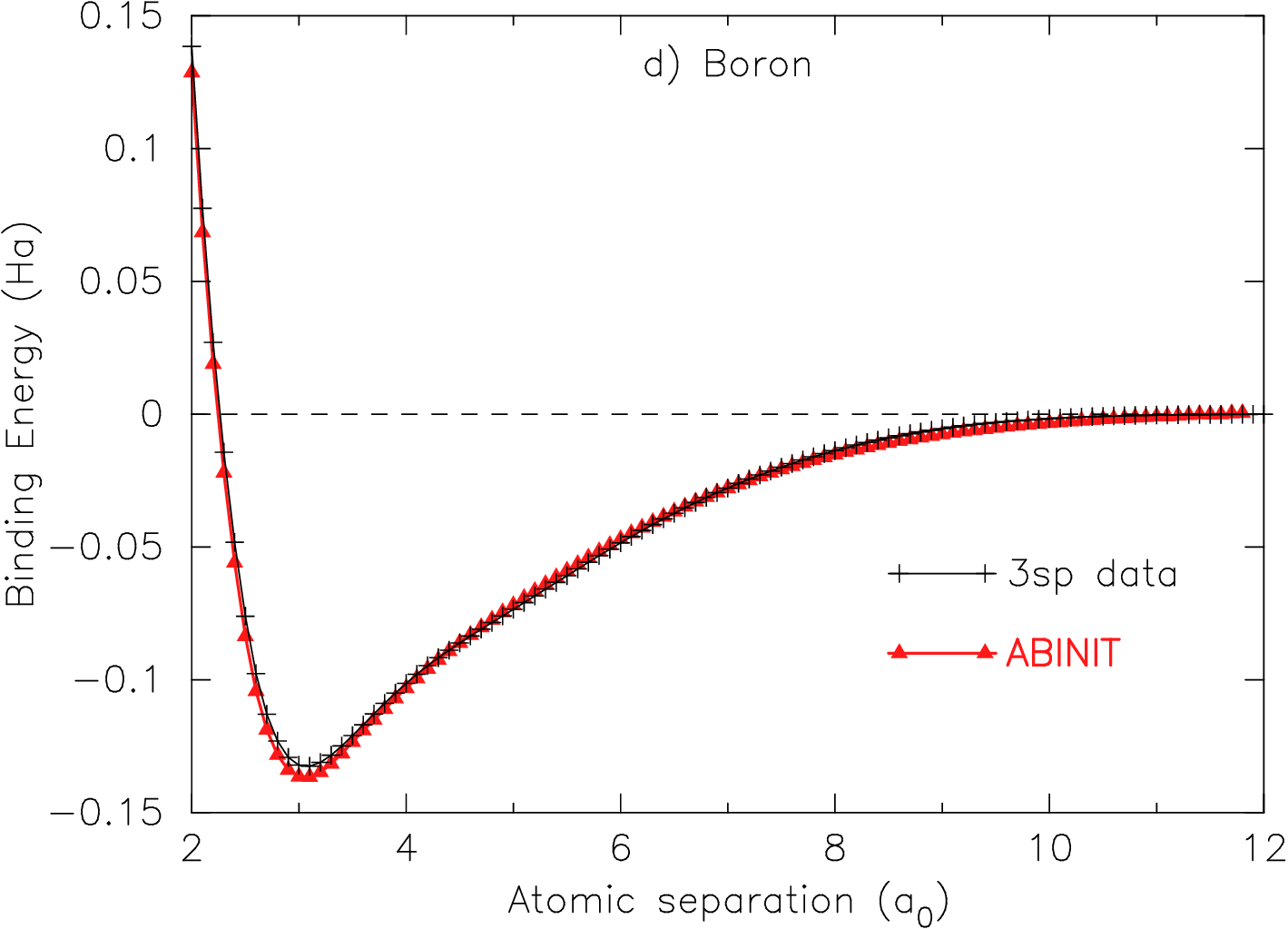}
  \includegraphics[angle=-90,width=7.4cm]{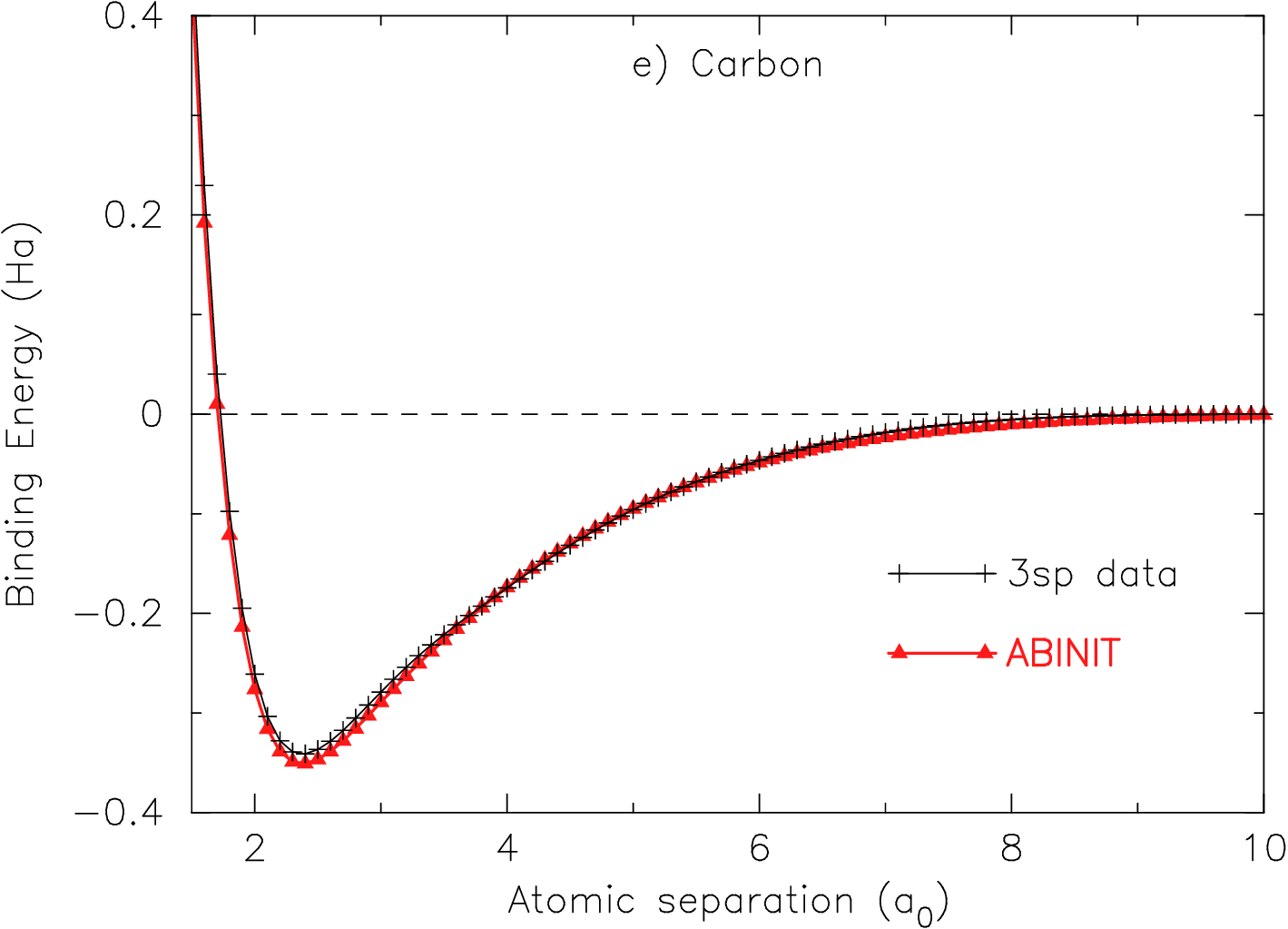}
  \includegraphics[angle=-90,width=7.4cm]{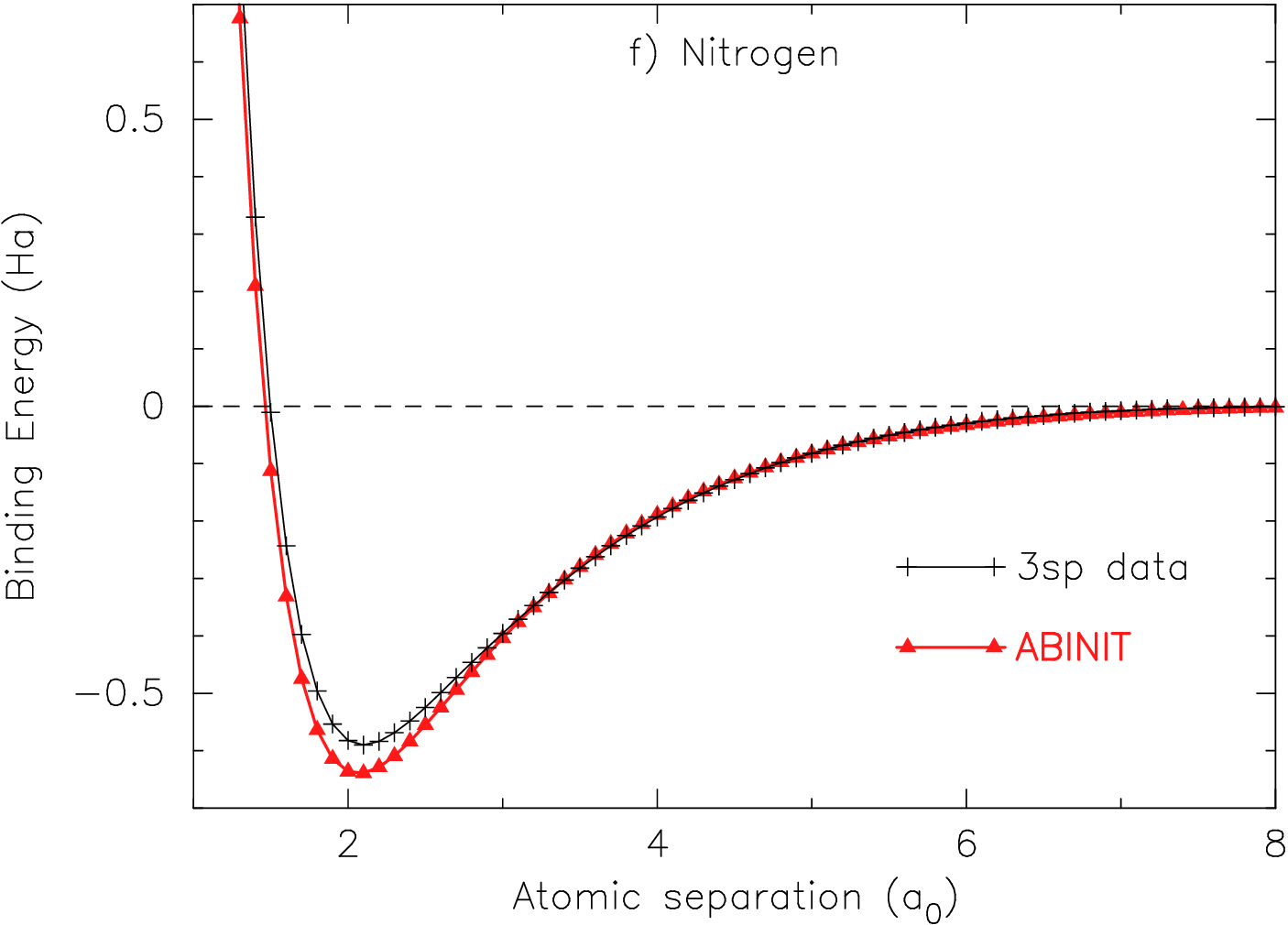}
  \includegraphics[angle=-90,width=7.4cm]{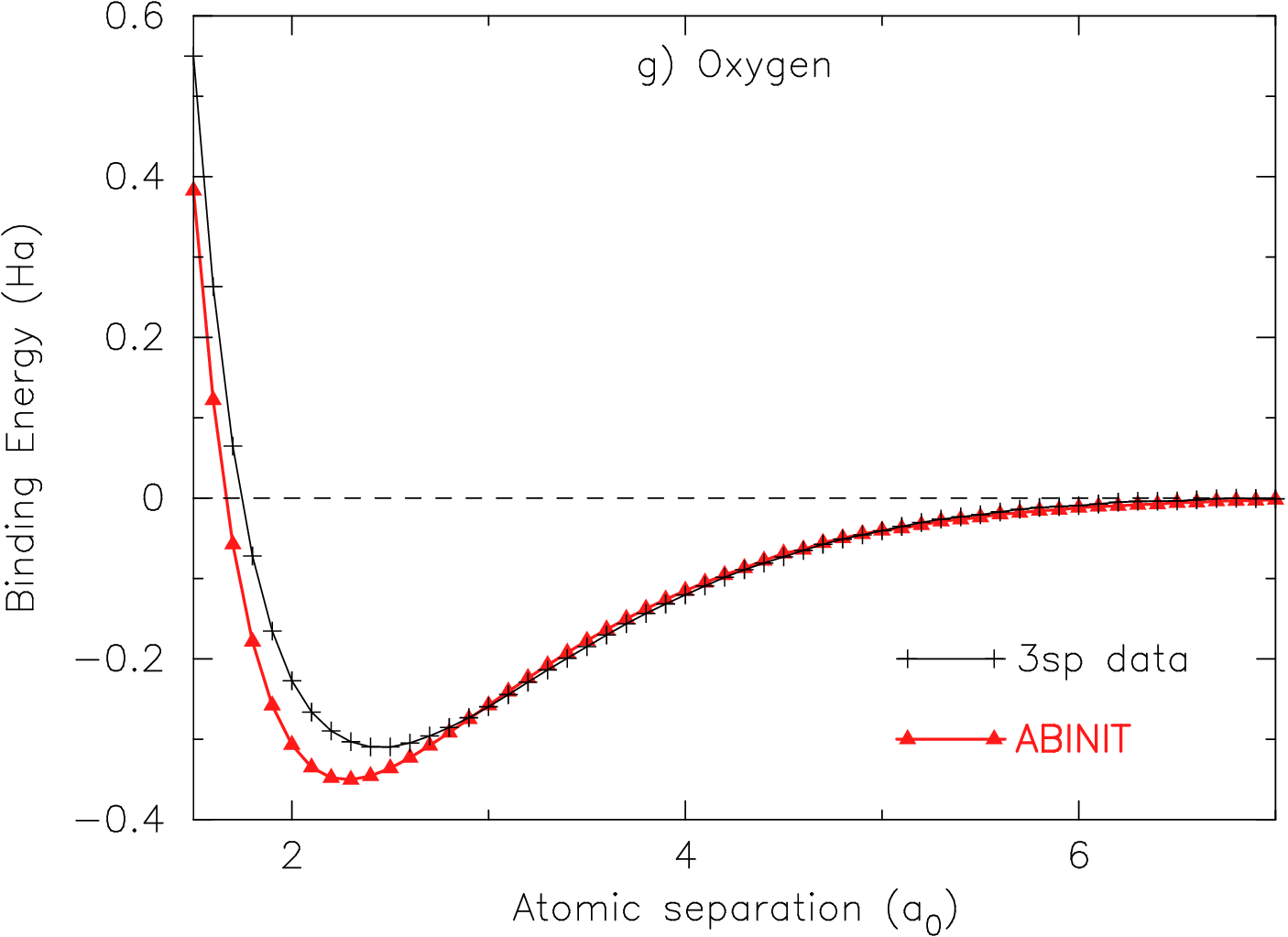}
  \includegraphics[angle=-90,width=7.4cm]{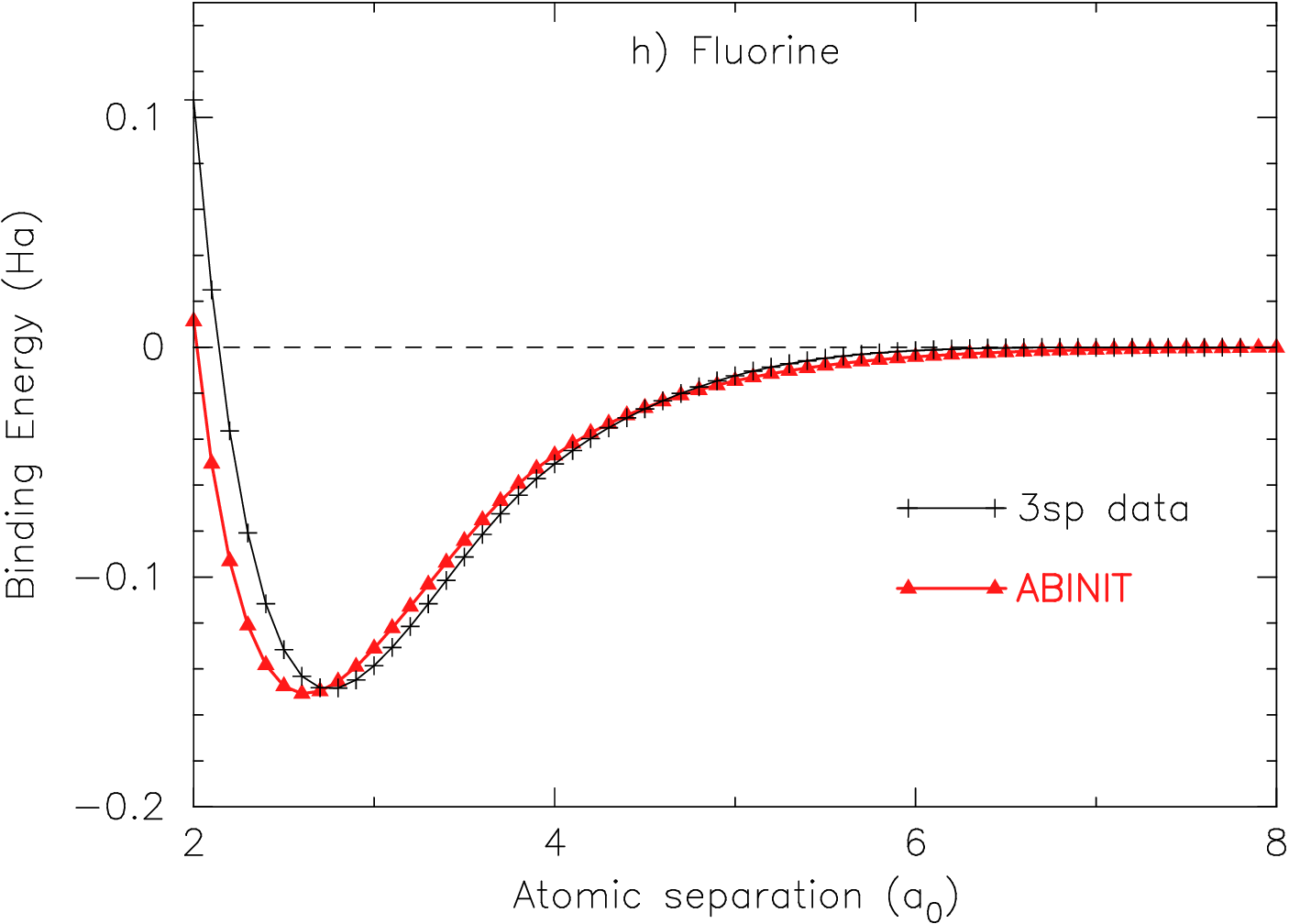}
  \caption{\label{fig_dimer}
  (Color online) The binding energy as a function of atomic distance 
  for the dimers (a) H$_2$ and (b)--(h) Li$_2$--F$_2$.
  The curves labeled \textsc{abinit} show reference results obtained with the computer
  program \textsc{abinit}.
  The curves labeled ``3\textit{sp} data'' show the results that we have obtained using
  optimized basis sets with $3$ different exponents, but identical for 
  \textit{s} and \textit{p} Gaussians.
  It should be stressed that the results for (b) Li and (c) Be were obtained using
  pseudopotentials of Hartwigsen \textit{et al.}, \cite{Hartwigsen98} whereas all other
  curves were calculated using Goedecker-Teter-Hutter pseudopotentials. \cite{Goedecker96}}
\end{figure*}

\begin{figure*}
  \includegraphics[angle=-90,width=7.4cm]{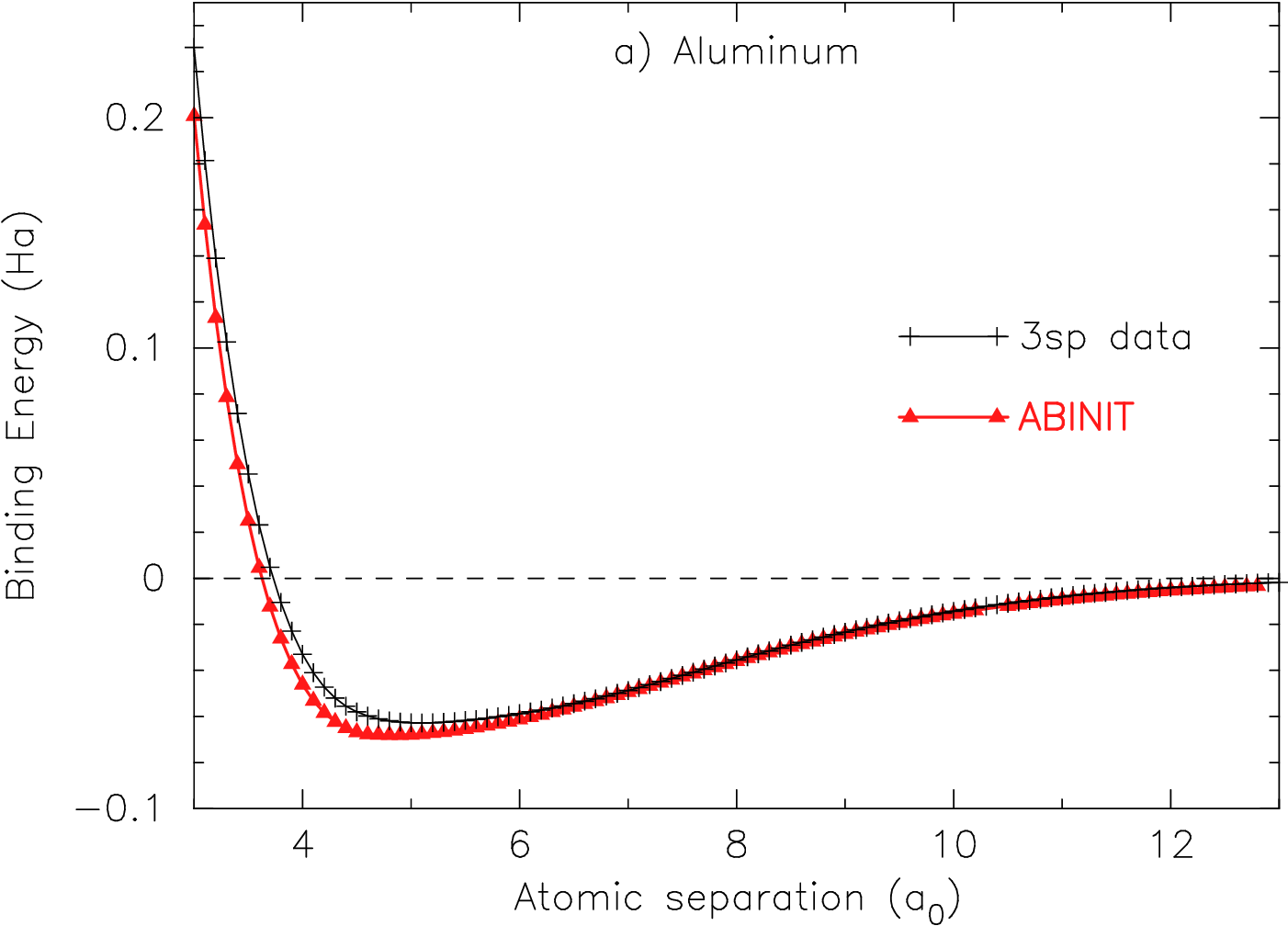}
  \includegraphics[angle=-90,width=7.4cm]{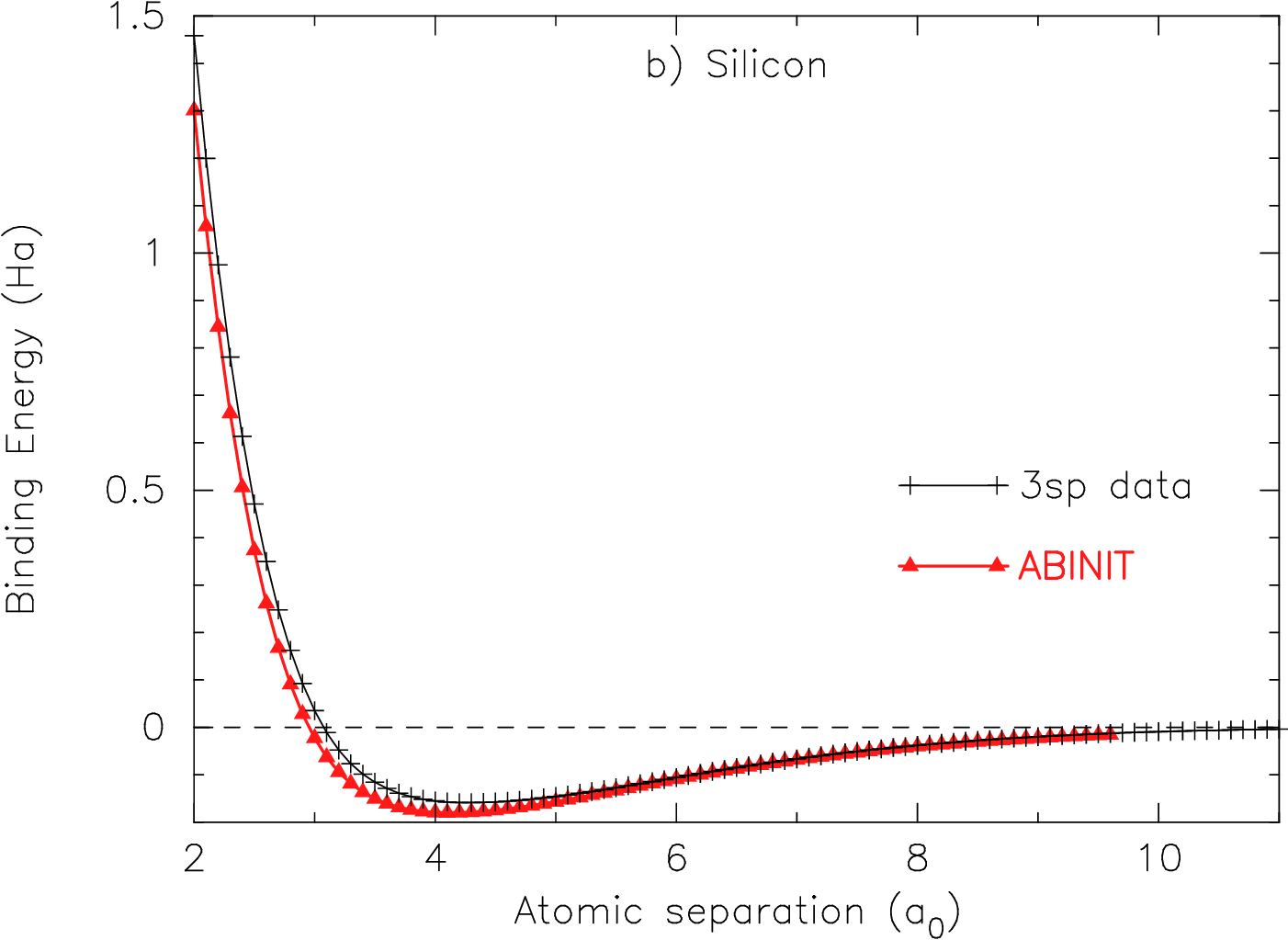}
  \includegraphics[angle=-90,width=7.4cm]{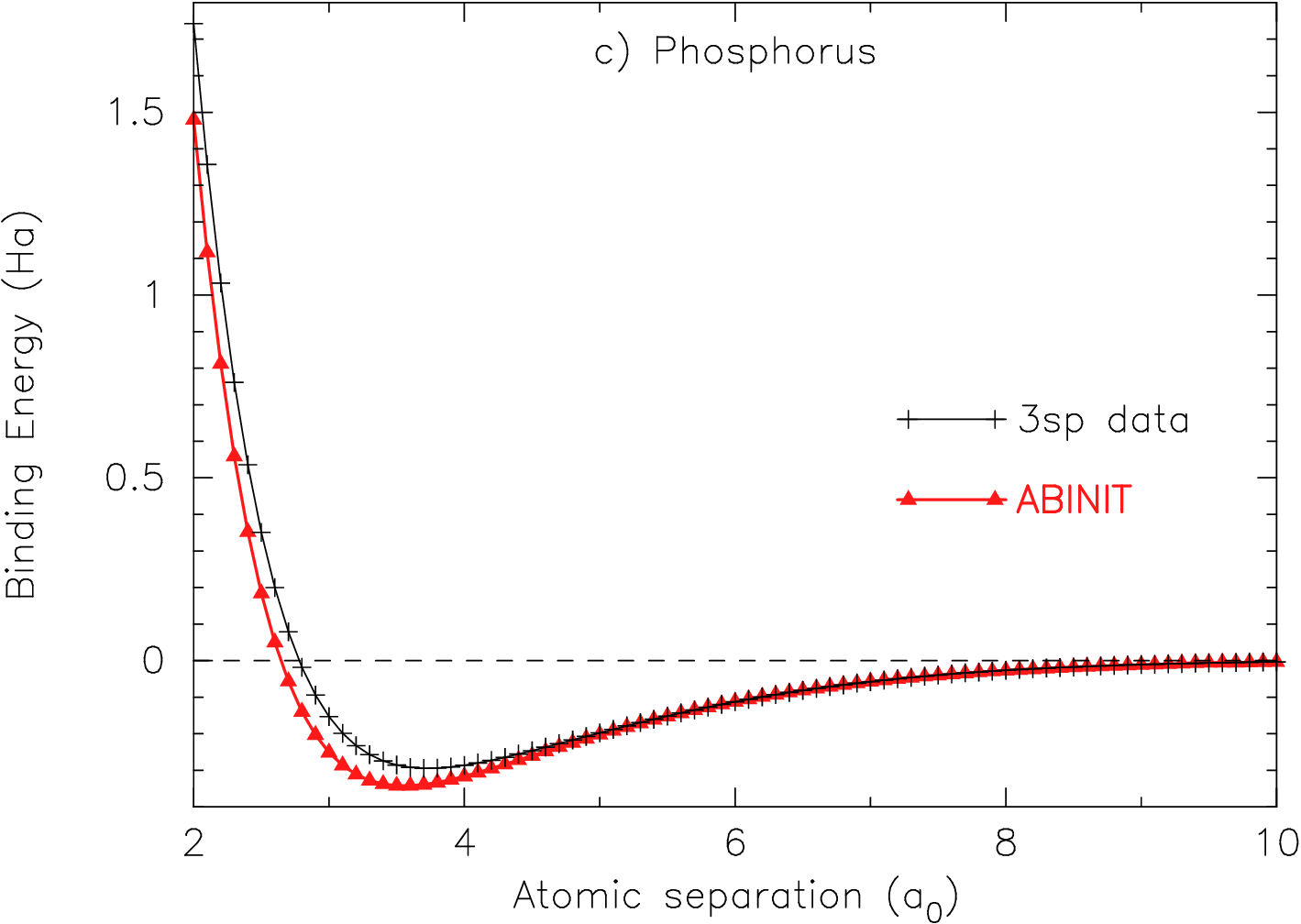}
  \includegraphics[angle=-90,width=7.4cm]{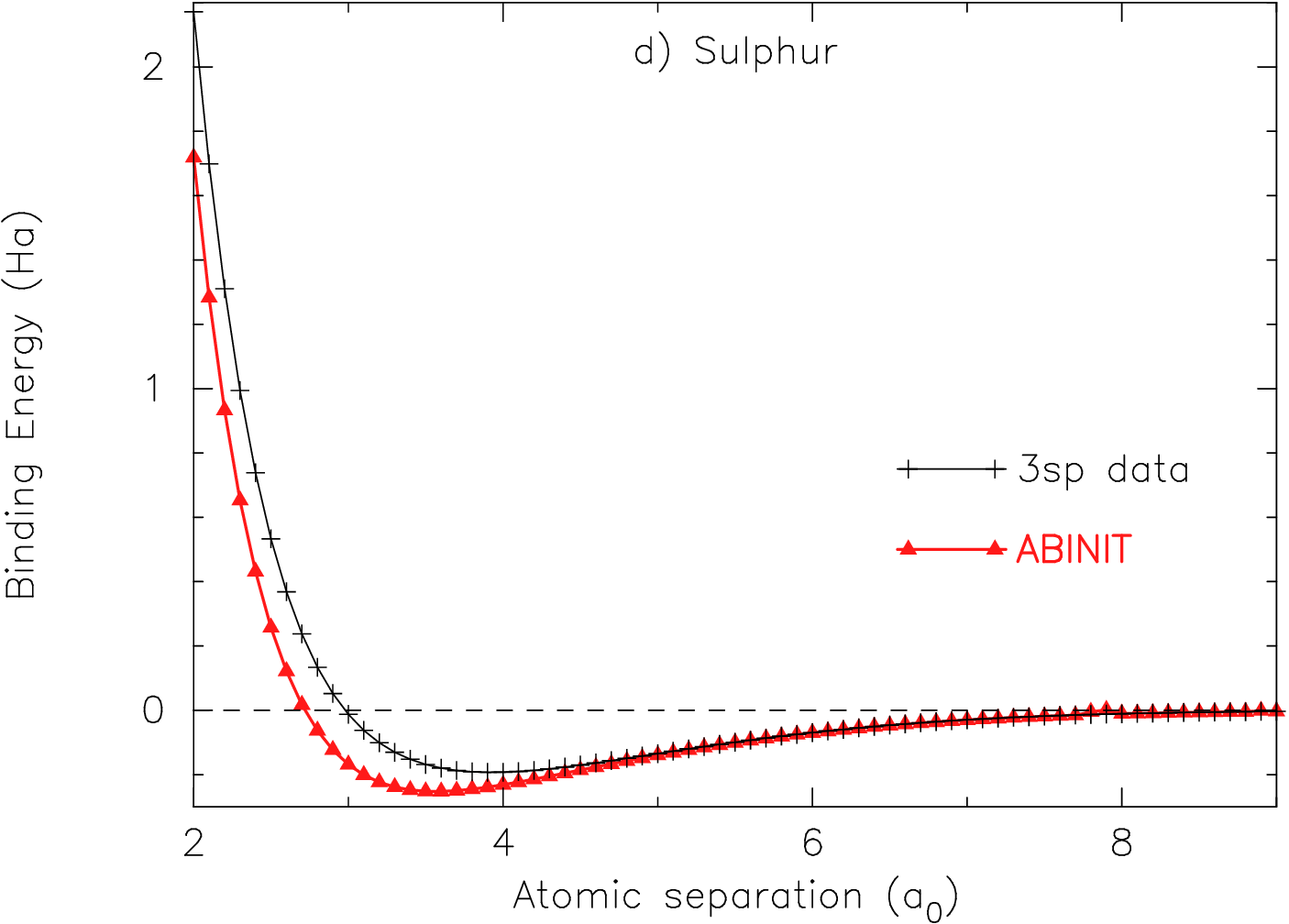}
  \includegraphics[angle=-90,width=7.4cm]{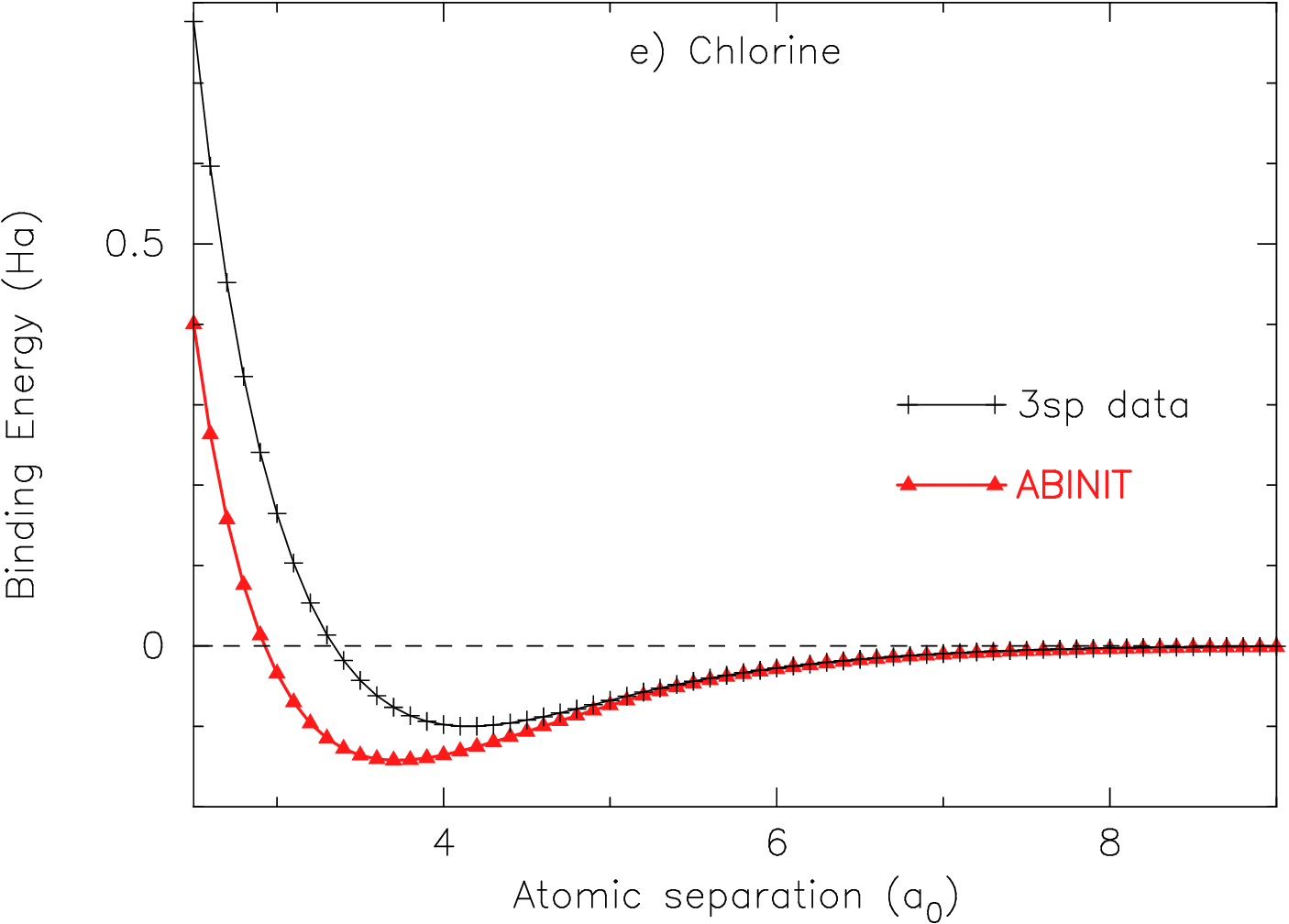}
  \caption{\label{fig_dimer2}
  (Color online) The binding energy as a function of atomic distance
  for dimers of the second-row elements Al--Cl.
  The curves labeled \textsc{abinit} show reference results obtained with the computer
  program \textsc{abinit}.
  The curves labeled ``3\textit{sp} data'' show the results that we have obtained using
  optimized basis sets with $3$ different exponents, but identical for
  \textit{s} and \textit{p} Gaussians.}
\end{figure*}

Surprisingly, we found that the 3\textit{sp} data gave close to optimal results for the
binding energies of all dimers.
The resulting curves are shown in Figs.\ \ref{fig_dimer} and \ref{fig_dimer2}.
For hydrogen, the 4\textit{sp} curve (not shown) 
was slightly better than the 3\textit{sp} curve, but the energy differences 
between both curves were small compared to the errors that we made in other
elements by restricting ourselves to a basis containing only \textit{s} and \textit{p} orbitals.
For lithium and beryllium the pseudopotentials of Goedecker, Teter, and Hutter \cite{Goedecker96}
treat the 1\textit{s} electrons as valence electrons, and offer therefore only a limited advantage
in speed as compared to all-electron calculations.
For this reason, for these two elements we decided to use the above-mentioned pseudopotentials
of Hartwigsen \textit{et al.}, \cite{Hartwigsen98} which for Li and Be happen to be parameterized by the same
analytical form as the pseudopotentials of Goedecker, Teter, and Hutter, so that no additional
programming was necessary for these elements.
The Li and Be pseudopotentials of Hartwigsen \textit{et al.} \cite{Hartwigsen98} that 
we have used treat the 1\textit{s} electrons as core
electrons, and should thus provide a considerable speed-up as compared to all-electron calculations.

As one can see in Fig.\ \ref{fig_dimer} our results for H, Li, Be, B, and C are excellent.
At first sight, the agreement of our dimer curve with the corresponding \textsc{abinit} reference curve
looks slightly worse for Be$_2$ [Fig.\ \ref{fig_dimer}(c)] than for the other dimers
[Figs.\ \ref{fig_dimer}(a), \ref{fig_dimer}(b), \ref{fig_dimer}(d), and \ref{fig_dimer}(e)].
The reason is that the binding energy of Be$_2$ at its equilibrium distance 
is approximately an order of magnitude less than that of, for example, H$_2$ or C$_2$.
As a result, the 3\textit{sp} basis set errors look large for Be$_2$ [Fig.\ \ref{fig_dimer}(c)], even though
the absolute error is less than $9$ mHa for all distances shown ($2.5$ mHa at the minimum), which should be acceptable for
most applications.
In Fig.\ \ref{fig_dimer},
starting from nitrogen, the incompleteness of our basis set becomes gradually noticed, and
as a result the agreement between our results and the \textsc{abinit} reference curves
becomes gradually worse.
The same behavior can be noticed in the dimer curves for the second-row elements, which are shown in
Fig.\ \ref{fig_dimer2}: Whereas the 3\textit{sp} basis set gives very reasonable results for Al, Si, and P,
the curves for sulphur and chlorine differ increasingly from the \textsc{abinit} reference curves.
It would, of course, depend on the application, whether the deviations are acceptable.
For future reference,
all our optimized 3\textit{sp} exponents for H, Li--F, and Al--Cl are presented in Table \ref{table_exps}.
\begin{table}
  \begin{tabular*}{0.45\textwidth}{@{\extracolsep{\fill}}ccccc}
  \hline
  Element & $d (a_0)$ & \multicolumn{3}{c}{Exponents $(a_0^{-2})$} \\
  \hline 
  H  & $1.45$  & $4.8336700$ & $0.7962986$ & $0.1703991$ \\
  Li & $5.099$ & $0.7107769$ & $0.1151854$ & $0.0381934$ \\
  Be & $4.515$ & $1.3420164$ & $0.2543896$ & $0.0760345$ \\
  B  & $3.00$  & $2.2292786$ & $0.4922699$ & $0.1321358$ \\
  C  & $2.35$  & $3.3582735$ & $0.7697578$ & $0.2000381$ \\
  N  & $2.07$  & $4.6545372$ & $1.0441223$ & $0.2624989$ \\
  O  & $2.27$  & $6.3165398$ & $1.4495363$ & $0.3579517$ \\
  F  & $2.61$  & $8.3059235$ & $1.9605670$ & $0.4596180$ \\
  Al & $4.64$  & $0.9496758$ & $0.2017059$ & $0.0659053$ \\
  Si & $4.29$  & $1.1985394$ & $0.2972473$ & $0.0941132$ \\
  P  & $3.57$  & $1.4859314$ & $0.4104534$ & $0.1277851$ \\
  S  & $3.57$  & $1.7834663$ & $0.5430840$ & $0.1651222$ \\
  Cl & $3.74$  & $2.1074139$ & $0.7324463$ & $0.2130882$ \\
  \hline
  \end{tabular*}
  \caption{\label{table_exps}
  Optimized Gaussian basis set exponents for H, Li--F, and Al--Cl.
  The same exponents should be used for Gaussian \textit{s} and \textit{p} orbitals. 
  The second column gives the atomic separations ($d$) of the dimers for which the exponents were optimized.
  It should be stressed that the exponents for Li and Be were optimized using
  pseudopotentials of Hartwigsen \textit{et al.}, \cite{Hartwigsen98} whereas all other
  exponents were optimized for Goedecker-Teter-Hutter pseudopotentials. \cite{Goedecker96}}
\end{table}

To test, how well the exponents of Table \ref{table_exps} perform in systems other than the dimers for which
they were optimized, we calculated the binding energies and the bond lengths for $8$ diatomic 
molecules.
Our results are given in Table \ref{table_diatom}.
\begin{table}
  \begin{tabular*}{0.45\textwidth}{@{\extracolsep{\fill}}ccccc}
  \hline
  Molecule & \multicolumn{2}{c}{Bond Length (a$_0$)} & \multicolumn{2}{c}{Binding Energy (Ha)} \\
           & $3$\textit{sp} data & \textsc{abinit} & $3$\textit{sp} data & \textsc{abinit} \\
  \hline
  BH  & $2.383$ & $2.366$ & $-0.1860$ & $-0.1857$ \\
  FH  & $1.784$ & $1.763$ & $-0.2970$ & $-0.3026$ \\
  AlH & $3.210$ & $3.161$ & $-0.1573$ & $-0.1599$ \\
  HCl & $2.451$ & $2.435$ & $-0.2212$ & $-0.2279$ \\
  CO  & $2.162$ & $2.127$ & $-0.5322$ & $-0.5715$ \\
  CS  & $3.003$ & $2.884$ & $-0.3422$ & $-0.3912$ \\
  PN  & $2.930$ & $2.789$ & $-0.3978$ & $-0.4528$ \\
  SiO & $2.977$ & $2.833$ & $-0.3854$ & $-0.4368$ \\
  \hline
  \end{tabular*}
  \caption{\label{table_diatom}
  Bond lengths and binding energies of diatomic molecules obtained using the $3$\textit{sp} Gaussian basis
  sets from Table \ref{table_exps}. 
  Reference results were calculated using the computer program \textsc{abinit}.}
\end{table}
The first thing to be noticed is that the bond lengths obtained with the Gaussian $3$\textit{sp} basis sets are always 
larger than the \textsc{abinit} reference bond lengths, which are also given in Table \ref{table_diatom}, and
that the $3$\textit{sp} binding energies are always smaller than the reference results,
with the exception of BH, for which the $3$\textit{sp} and the reference binding energies agree almost perfectly.
These trends are consistent with what can be observed
for the dimers (Figs.\ \ref{fig_dimer} and \ref{fig_dimer2}).
A more quantitative analysis shows that
for BH, FH, AlH, HCl, and CO the Gaussian $3$\textit{sp} basis reproduces the \textsc{abinit} bond lengths to within $2\%$,
and that for CS, PN, and SiO the deviations of the bond lengths are of the order of $5\%$.
The errors in the binding energies are between $0.3$ and $6.7$ mHa for BH, FH, AlH, and ClH, and between
$39.3$ and $55$ mHa for CO, CS, PN, and SiO, but never more than $12.5\%$ of the \textsc{abinit} reference binding energy.
All these errors lie within the range of what one might reasonably have expected from the dimer curves 
(Figs.\ \ref{fig_dimer} and \ref{fig_dimer2}), and we conclude that our Gaussian $3$\textit{sp} basis sets
perform equally well for the dimers and the small molecules that we have studied.
We once again remind the reader that our main goal in the present work is to compete with
tight-binding theory rather than to produce very accurate results for a variety of small molecules
composed of first and second row
elements.

\section{\label{sec_conclusion}Conclusion}

In this paper we have optimized Gaussian exponents for the Goedecker-Teter-Hutter pseudopotentials \cite{Goedecker96}
for all elements in the first two rows of the periodic table, with the exception of the noble gases and Na and Mg.
We decided not to include the noble gases in our study, because the local density approximation \cite{Perdew92}
is known to perform quite badly for these elements.
Sodium and magnesium were kept out of the scope of this study, because the Goedecker-Teter-Hutter pseudopotentials
\cite{Goedecker96} treat the 2\textit{s} and 2\textit{p} electrons as valence electrons,
which would make practical applications unnecessarily expensive in terms of computer time.

The main conclusion of the present work is
that for all elements studied it is sufficient to use a Gaussian basis containing only 
\textit{s} and \textit{p} orbitals, and furthermore that only three Gaussian exponents, which can be chosen equal for the
\textit{s} and the \textit{p} channels, are needed to obtain a reasonable accuracy, as has been discussed in the
body of this paper.

We believe that our results constitute an important contribution towards the development of fast DFT
computer programs using both pseudopotentials and Gaussian basis sets.

\acknowledgments

This work has been supported by the Deutsche Forschungsgemeinschaft (DFG)
through the priority program SPP 1134 and by the European Community Research
Training Network FLASH (MRTN-CT-2003- 503641).


\begin{thebibliography}{11}
\bibitem[{foo()}]{footnote1}
\bibinfo{note}{In tight-binding calculations studying phase transitions induced
  by femtosecond laser pulses good tight-binding parameters have been found for
  C and Si, but not for other elements.}
\end{thebibliography}

\end{document}